\documentclass[twocolumn]{aastex63}
\usepackage{amssymb,amsmath}
\usepackage{graphicx}
\usepackage{xcolor,natbib}
\usepackage{multirow}
\usepackage[T1]{fontenc}
\usepackage{ae,aecompl}
\usepackage{newtxtext, newtxmath}
\usepackage{float}
\usepackage{subfigure}
\usepackage{longtable}
\usepackage{enumitem}
\usepackage{longtable,listings}
\usepackage[flushleft]{threeparttable}
\usepackage{parcolumns}
\def\oiii{[O~{\sc iii}]\ }
\def\nii{[N~{\sc ii}]\ }
\def\sii{[S~{\sc ii}]$\lambda6717,6731$\AA\ }
\def\oii{[O~{\sc ii}]$\lambda3727$\AA\ }

\submitjournal{ApJ}

\shorttitle{broad Blue-shifted \oiii components in quasars}
\shortauthors{ZHANG}

\begin{document}

\title{On strong correlation between shifted velocity and line width of broad blue-shifted \oiii components in quasars}

\correspondingauthor{XueGuang Zhang}
\author{XueGuang Zhang}
\affiliation{School of Physics and technology, Nanjing Normal University, No. 1, Wenyuan Road, Nanjing, 210023, P. R. China}

\begin{abstract}
    In this manuscript, we report strong linear correlation between shifted velocity and line width of the broad 
blue-shifted \oiii components in SDSS quasars. Broad blue-shifted \oiii components are commonly treated as indicators 
of outflows related to central engine, however, it is still an open question whether the outflows are related to 
central accretion properties or related to local physical properties of NLRs (narrow emission line regions). 
Here, the reported strong linear correlation with the Spearman Rank correlation coefficient 0.75 can be expected 
under the assumption of AGN (active galactic nuclei) feedback driven outflows, through a large sample of 
535 SDSS quasars with reliable blue-shifted broad \oiii components. Moreover, there are much different detection 
rates for broad blue-shifted and broad red-shifted \oiii components in quasars, and no positive correlation can 
be found between shifted velocity and line width of the broad red-shifted \oiii components, which provide further 
and strong evidence to reject possibility of local outflows in NLRs leading to the broad blue-shifted \oiii 
components in quasars. Thus, the strong linear correlation can be treated as strong evidence for the broad 
blue-shifted \oiii components as better indicators of outflows related to central engine in AGN. Furthermore, 
rather than central BH masses, Eddington ratios and continuum luminosities have key roles on properties of the 
broad blue-shifted \oiii components in quasars. 
\end{abstract}


\section{Introduction}

   Active Galactic Nuclei (AGN) driven outflows, as the probe of AGN feedback, have been studied in detail 
for more than two decades \citep{ck03, vc05, fa12, ps12, kp15, cb16, mb18}. AGN feedback driven outflows not 
only show clear linkages between AGN and host galaxies, such as AGN feedback model expected M-sigma relations 
\citep{fm00, geb00, kh13} indicating mass outflowing leading to the strong physical connections between central 
black hole (BH) masses and host galaxy properties, but also strongly indicate outflows from central regions 
have apparent and important effects on structures of emission/absorption lines from both NLRs (narrow emission 
line regions) and BLRs (broad emission line regions).

   Blue-shifted \oiii emission features have been widely treated as indicators of outflows on scale of kpcs 
around NLRs in AGN, besides broad absorption lines in UV and X-ray bands as better indicators of outflows coming 
from central accretion disk winds in AGN on scale of light-months to light-years (around central BLRs) \citep{gb07, 
tm15}. \citet{sp97} have shown that broad wings of \oiii emission lines could be emitted from the outer BLRs and 
suggested the presence of an outflow component, through study of variations of the \oiii line profiles in NGC5548. 
\citet{tw01} have shown that kinematics of the broad blue-shifted \oiii emission lines are consistent with outflow 
in an inner NLRs in PKS1549-79. \citet{gs05} have shown that kinematic properties of the blue-shifted absorption-line 
system (relative to the emission-line system) are similar to the blue-shifted \oiii lines, indicating strong 
connections with outflowing materials in 3C48. \citet{ht08} have discussed fast outflows in compact radio sources, 
through properties of broad blue-shifted \oiii components. \citet{ma13} have shown that the \oiii profiles of type-1 
and type-2 AGN show the same trends in terms of line width, but type-1 AGN display a much stronger blue wings, 
which can be well interpret as evidence of outflowing ionized gases. \citet{pb15} have detected galaxy-wide outflows 
in high redshift luminous obscured quasars by properties of broad \oiii lines. \citet{za16} have shown that broad 
\oiii emission regions on a few kpc scales can be affected by extreme outflow from central regions through a sample 
of high redshift red quasars. However, not similar as broad absorption lines tightly related to outflows from 
central disk winds, local physical properties in NLRs can also lead to blue-shifted features in \oiii emission 
lines, such as local flows in NLRs related to stellar winds. Certainly, there are some reported results indicating 
blue-shifted \oiii components could be possibly related to central engine. \citet{tb05} have shown weak dependence 
of shifted velocities of blue-shifted \oiii components on line width of \oiii lines through a sample of about 800 
quasars in SDSS DR1 (Sloan Digital Sky Survey, Data Release 1) \citep{ab03}. \citet{so18} have shown much loose 
correlation between shifted velocities of blue-shifted \oiii components and line widths of \oiii lines through a 
sample of 28 narrow line Seyfert I galaxies, much similar weak trends can also be found in \citet{wo16, eu17} 
in samples of type 2 AGN and hidden type-1 AGN.

    More and more evidence have shown that broad blue-shifted \oiii components rather than the core \oiii 
components are more tightly related to central engine in AGN, such as our previous results in \citet{zh17} and 
previous results in \citet{za16}, strongly indicating that broad \oiii emission regions are very nearer to central 
regions and have stronger luminosity dependence on central continuum emissions in AGN. In the manuscript, 
unless otherwise stated, the broad \oiii components mean the emissions from the non-BLR regions. Thus, it is 
interesting to check whether are there apparently stronger evidence to support the broad blue-shifted \oiii 
emission components, rather than the blue-shifted asymmetric properties determined from the full \oiii as 
discussed in the literature, treated as indicators of outflows from central regions in AGN. Certainly, after 
considering probably serious obscuration on broad \oiii emissions in type-2 AGN, we will do our study through 
a sample of broad line AGN. And the manuscript is organized as follows. In section 2, we show our sample 
selection and emission line fitting procedure. In Section 3, we show our main results on properties of broad 
\oiii components in a large sample of SDSS quasars in DR12 (Data Release 12, \citet{al15}). In section 4, the 
main discussions are given. Then, in Section 5, we show our main conclusions. And in this manuscript, 
we have adopted the cosmological parameters of $H_{0}=70{\rm km\cdot s}^{-1}{\rm Mpc}^{-1}$, 
$\Omega_{\Lambda}=0.7$ and $\Omega_{\rm m}=0.3$.

\begin{figure*}
\centering\includegraphics[width = 18cm,height=6cm]{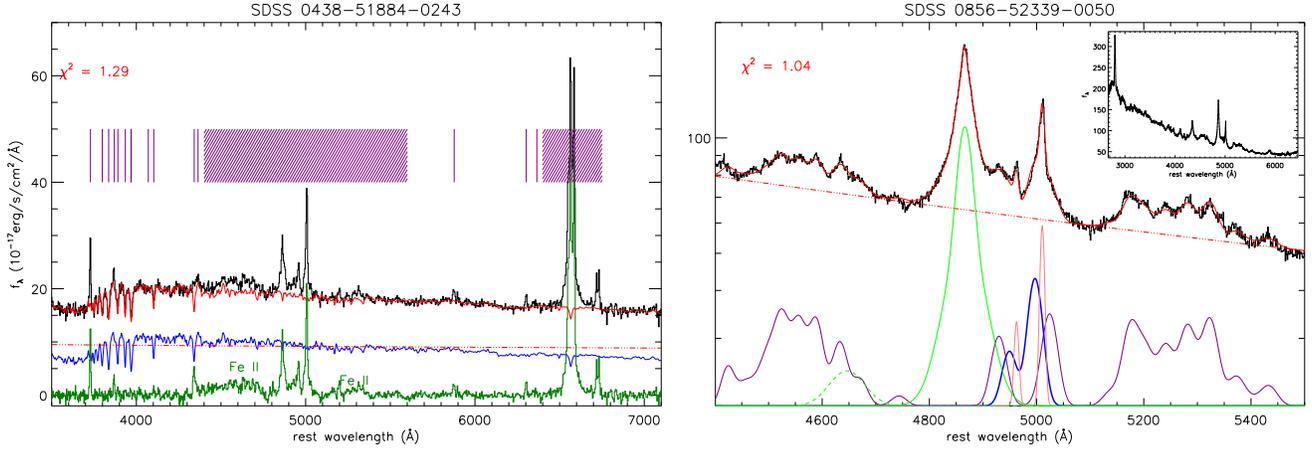}
\caption{Left panel shows the SSP method determined stellar lights in the quasar SDSS 0438-51884-0243. Right 
panel shows the best fitted results to the emission lines around H$\beta$ in the quasar SDSS 0856-52339-0050. 
In left panel, from top to bottom, solid lines in black and in red show the observed SDSS spectrum and the best 
fitted results by broadened stellar templates plus a power law component, double-dot-dashed red line shows 
the determined power law AGN continuum emissions, solid blue line shows the determined stellar lights, solid green 
line shows the pure line spectrum after subtractions of both stellar lights and the AGN continuum emissions. 
When the SSP method is applied, the emission lines masked out are marked by the vertical lines in purple and 
in the two areas filled by purple lines. From left to right, the vertical lines in purple mark the following 
emission features masked out, including \oii, H$\theta$, H$\eta$, [Ne~{\sc iii}]$\lambda3869$\AA, 
He~{\sc i}$\lambda3891$\AA, Calcium K line, [Ne~{\sc iii}]$\lambda3968$\AA, Calcium H line, 
[S~{\sc ii}]$\lambda4070$\AA, H$\delta$, H$\gamma$, [O~{\sc iii}]$\lambda4364$\AA, He~{\sc i}$\lambda5877$\AA\ 
and [O~{\sc i}]$\lambda6300,6363$\AA, respectively. The area filled by purple lines around 5000\AA\ shows the 
region masked out including the emission features of optical Fe~{\sc ii} lines, broad and narrow H$\beta$ and 
\oiii doublet, and the area filled by purple lines around 6550\AA\ shows the region masked out including the 
emission features of broad and narrow H$\alpha$, \nii and \sii doublets. The determined $\chi^2$ value of 1.29 
for the best fitted results is marked in the top-left corner in left panel. In the right panel, from top to 
bottom, solid black line shows the observed line spectrum, solid red line shows the determined best fitted 
results, double-dot-dashed red line shows the determined power law continuum emissions, solid green line 
shows the determined broad H$\beta$, solid purple line shows the determined optical Fe~{\sc ii} lines, dashed 
green line shows the determined broad He~{\sc ii} line, solid pink line shows the determined core \oiii 
components, and thick blue solid line shows the determined broad blue-shifted \oiii components. The top-right 
corner shows the observed blue spectrum of SDSS 0856-52339-0050, in order to clearly show there are few 
contributions of stellar lights in the spectrum. And the calculated $\chi^2$ value of 1.04 for the best fitted 
results is marked in the top-left corner in the right panel.}
\label{line}
\end{figure*}

\begin{figure*}
	\centering\includegraphics[width = 18cm,height=6cm]{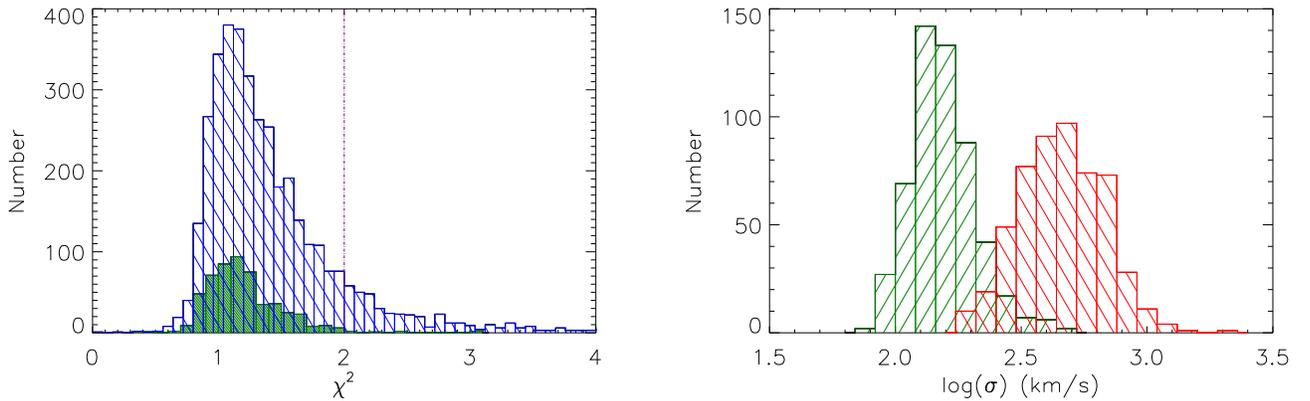}
	\caption{Left panel shows the distribution of $\chi^2$ for the best fitted results to the emission lines in all
	the collected 3735 quasars (histogram in blue) and in the 535 quasars with reliable broad \oiii components
	(histogram in dark green). Vertical red line shows the position of $\chi^2=2$. Right panel shows the distributions
	of line width of the broad (histogram in red) and the core (histogram in dark green) \oiii components of the 535 quasars.
	}
	\label{dis2}
\end{figure*}

\section{Sample Selection and Emission line Fitting procedure}

    The work is based on a large sample of quasars with broad shifted \oiii emission components relative 
to narrow core \oiii components. So that, there are three steps to create our main sample. The first step 
is to create a parent sample of quasars. The second step is to measure necessary line parameters, especially 
parameters of \oiii emission lines. And the third step is to create our final main sample including objects 
with reliable broad blue-shifted \oiii emission components, based on reliable measured parameters.

    SDSS SkyServer provided SQL (Structured Query Language) Search tool 
(\url{http://skyserver.sdss.org/dr12/en/tools/search/sql.aspx}) is firstly applied to conveniently collect 
SDSS quasars \citep{rg02, ro12} from SDSS DR12 to create the parent sample. The applied query is as follows:
\begin{lstlisting}
SELECT  plate, fiberid, mjd  
FROM    SpecObjall
WHERE   class = 'QSO' and z < 0.65 and 
        zwarning = 0 and snmedian > 20
\end{lstlisting},
where "SpecObjall" represents the SDSS provided dataset including basic information (redshift, classification, 
signal-to-noise (SN) of spectra, etc.) of all objects in SDSS DR12. The query leads to collection of 3735 SDSS 
quasars with reliable redshift less than 0.65 and with high quality spectra with median SN larger than 20. The 
criterion of $z<0.65$ is applied to ensure \oiii emission lines totally covered in SDSS spectra. The 
criterion of median $SN>20$ on the full observed spectrum can lead to more neat and clean spectroscopic 
emission line features.

    The second step is to measure necessary emission line parameters for the quasars in the parent sample. 
Similar as what we have done in \citet{zh14, zh16, ra17, zh19}, the most commonly accepted SSP method 
\citep{bc03, ka03, cm05} has been firstly applied to a small number of the collected SDSS quasars (especially 
the low redshift quasars with $z<0.35$) of which spectra probably include apparent contributions of stellar 
lights, by considering broadened stellar templates plus a power law component. The 39 simple stellar population 
templates from \citet{bc03} have been exploited with the population ages from 5 Myr to 12 Gyr and with three 
solar metallicities (Z = 0.008, 0.05, 0.02), which can be used to well-describe the characteristics of almost 
all the SDSS galaxies as detailed discussions in \citet{bc03}. Through the Levenberg-Marquardt least-squares 
minimization technique (the MPFIT procedure) applied to the SDSS spectra with the emission lines being masked 
out, the stellar velocity dispersions can be determined by the broadened velocities to the stellar templates, 
and then the stellar component and the AGN power law continuum component can be clearly determined and separated. 
Here, we have not only masked out all the listed 24 narrow emissions lines with rest central wavelengths larger 
than 3700\AA\ in \url{http://classic.sdss.org/dr1/algorithms/speclinefits.html#linelist} with widths of about 
${\rm 450km/s}$, mainly including the [O~{\sc ii}]$\lambda3727$\AA, narrow H$\alpha$, narrow H$\beta$, narrow 
H$\gamma$, narrow H$\delta$, [O~{\sc iii}]$\lambda4364$\AA, [O~{\sc iii}]$\lambda4959,5007$\AA, 
[O~{\sc i}]$\lambda6300, 6363$\AA, [N~{\sc ii}]$\lambda6548,6583$\AA\ and [S~{\sc ii}]$\lambda6716, 6731$\AA, 
etc., but also masked out the optical Fe~{\sc ii} lines and the broad H$\alpha$ and H$\beta$. Based on the model 
determined stellar velocity dispersions and uncertainties (the returned best-fit parameters and the returned 
PERROR in the MPFIT procedure), the criterion that stellar velocity dispersions larger than 70${\rm km/s}$ and 
smaller than ${\rm 350km/s}$ at least 5 times larger than their corresponding uncertainties have been applied 
to determine that the determined stellar components are reliable enough. Here, we do not show further discussions 
on the SSP method, but the left panel of Fig.~\ref{line} shows an example on the SSP method determined stellar 
lights in the SDSS quasar PLATE-MJD-FIBERID=0438-51884-0243.

   After necessary subtractions of contributions of stellar lights, emission lines in line spectrum can be 
well described. Here, we mainly consider emission lines around H$\beta$ with rest wavelength range from 
4400\AA\ to 5600\AA, including broad H$\beta$, narrow H$\beta$, core and broad \oiii components, broad He~{\sc ii} 
and broad optical Fe~{\sc ii} lines. The emission features are fitted simultaneously by the following model 
functions. There are two (or more if necessary, after checking fitted results) broad Gaussian functions 
$G_{\rm B1}+G_{\rm B2}$ applied to describe broad H$\beta$. Here, each Gaussian function includes three 
parameters of central wavelength, second moment (width of the component) and line flux. And we accepted 
the central wavelengths of the two broad H$\beta$ components in the range of 4800\AA~ to 4900\AA. There are 
three narrow Gaussian functions $G_{\rm NH}+G_{\rm CO31}+G_{\rm CO32}$ applied to describe the narrow 
H$\beta$ and core \oiii components, with the central wavelength of narrow H$\beta$ in the range of 4840\AA~ 
to 4880\AA, and with the flux ratio of core \oiii component tied to the theoretical value of 3 \citep{dk07}, 
and with the central wavelengths of the three narrow components tied to be 4862.81\AA:4960.295\AA:5008.24\AA, 
and with the core \oiii components to have the same line width. There are two another Gaussian functions 
$G_{\rm BO31}+G_{\rm BO32}$ applied to describe the broad \oiii components, with the central wavelength 
of the broad [O~{\sc iii}]$\lambda5007$\AA\ in the range of 4980\AA~ to 5030\AA, and with the broad \oiii 
components to have the same redshift, the same line width, and to have the flux ratio tied to the theoretical 
value of 3. There is one broad Gaussian function $G_{\rm HeII}$  applied to describe weak He~{\sc ii} line, 
with the central wavelength in the range of 4600\AA~ to 4730\AA. There is one power law function 
$P_\lambda=\alpha\times(\lambda/5100\text{\AA})^\beta$ applied to describe AGN continuum emissions.
The broadened and scaled Fe~{\sc ii} templates discussed in \citet{kp10} $Fe_{\rm temps}$ is applied to 
describe probable optical Fe~{\sc ii} lines.  Finally, the detailed model functions are
\begin{equation}
\begin{split}
Y_{\rm model}&=G_{\rm B1}+G_{\rm B2}+G_{\rm NH}+G_{\rm CO31}+G_{\rm CO32}\\
&+G_{\rm BO31}+G_{\rm BO32}+G_{\rm HeII}+P_\lambda+Fe_{\rm temps}
\end{split}
\end{equation}.
Based on the widely applied Levenberg-Marquardt least-squares minimization technique, the best fitted results 
to the emission lines can be well determined, and the line parameters and corresponding uncertainties can also be 
well determined. Here, the uncertainties are the formal $1\sigma$ errors computed from the covariance matrix for the 
final determined best-fit model parameters returned by MPFIT procedure (\url{http://cow.physics.wisc.edu/~craigm/idl/idl.html}). 
Then, for the objects with the calculated $\chi^2=SSR/Dof>\sim2$ (where $SSR$ and $Dof$ as summed squared residuals 
and degree of freedom, respectively)\footnote{Among the collected 3735 SDSS quasars, there are 447 quasars with 
the best fitted results leading to $\chi^2$ larger than 2.}, the best fitted results have been carefully re-checked 
by eyes, in order to determine whether the fitting procedure should be re-applied with more than two broad components 
to describe the broad H$\beta$. Right panel of Fig.~\ref{line} shows one example on the best fitted results to 
the emission lines in the quasar SDSS 0856-52339-0050. And the distribution of final determined $\chi^2$ has 
been shown in the left panel of Fig.~\ref{dis2} for all the 3735 quasars.

\begin{figure*}
\centering\includegraphics[width = 18cm,height=6cm]{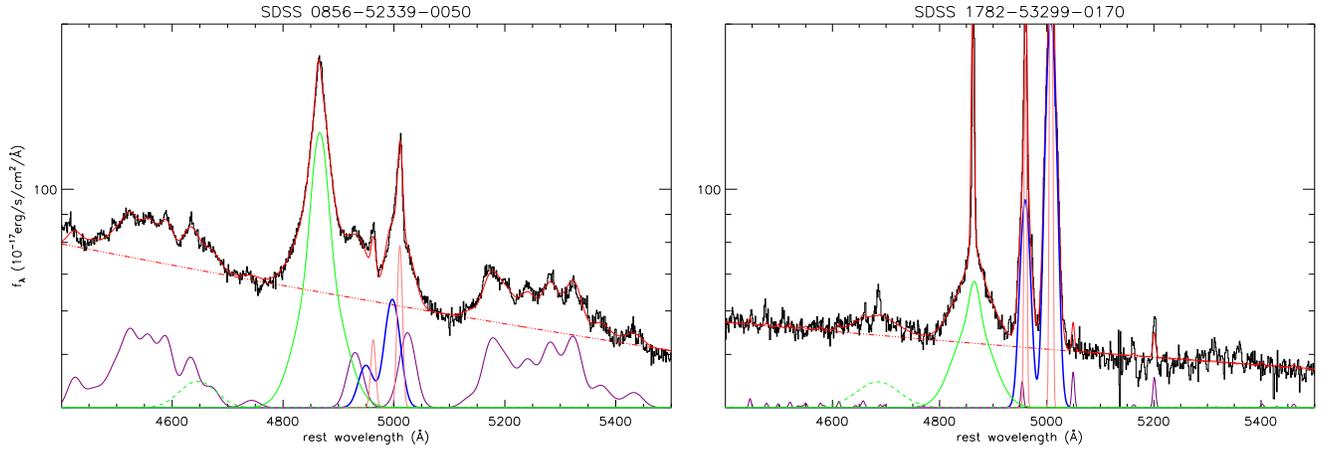}
\caption{Two examples on the MEM determined best fitted results to the emission lines around H$\beta$ 
in SDSS 0856-52339-0050 (left panel) and in SDSS 1782-53299-0170 (right panel). In each panel, line styles 
have the same meanings as those shown in the right panel of Fig.~\ref{line}, but the solid red line shows 
the MEM determined best fitted results.
}
\label{mcmc}
\end{figure*}

   Furthermore, in order to find more robust uncertainties of the line parameters, the commonly applied 
Maximum Likelihood Method (MEM) through the MCMC (Markov Chain Monte Carlo) technique \citep{fh13} has 
been accepted to re-fit the emission lines with the starting values of model parameters to be the values 
determined by the MPFIT procedure. Then, the MCMC technique determined posterior distribution can provide 
optional uncertainty of each model parameter. Here, as an two examples, Fig.~\ref{mcmc} shows the MEM 
determined best fitted results (which are totally similar as the MPFIT procedure determined best fitted results) 
to the emission lines around H$\beta$ in SDSS 0856-52339-0050 and in SDSS 1782-53299-0170. For the 
broad \oiii component in SDSS 0856-52339-0050, the central rest wavelength and the line width can be 
determined as $4997.09\pm0.40$\AA\ and $781\pm25{\rm km/s}$ based on the MCMC technique determined posterior 
distributions. Meanwhile, the MPFIT procedure determined rest central wavelength and line width of the broad 
\oiii component are $4997.06\pm0.84$\AA\ and $789\pm27{\rm km/s}$. There are similar uncertainties in the 
line width by the MCMC technique and by the MPFIT procedure, but smaller uncertainties in the rest 
central wavelength by the MCMC technique. However, for the broad \oiii component in SDSS 1782-53299-0170, 
the determined rest central wavelength and line width of the broad \oiii component are ($5007.73\pm0.05$\AA~ 
and $5007.85\pm0.23$\AA) and ($536\pm4{\rm km/s}$ and $535\pm15{\rm km/s}$) through the MPFIT procedure and 
through the MCMC technique, respectively, the uncertainties through the MCMC technique are about 4 times 
larger than the uncertainties determined by the MPFIT procedure. Therefore, different techniques can lead 
to different uncertainties of the model parameters. And in the manuscript, between the uncertainties determined 
by the MCMC technique and by the MPFIT procedure for each model parameter, the larger uncertainty has been accepted 
as its final uncertainty.

    Based on the measured line parameters and corresponding uncertainties of the core and the broad \oiii 
components, the broader component is accepted as the broad \oiii component, and the narrower component is 
accepted as the core \oiii component, and the shifted velocities of broad \oiii components relative to the 
core \oiii components can be calculated by $\Delta V=\lambda_{\rm 0,~core}-\lambda_{\rm 0,~broad}$, with 
the uncertainties of $\Delta V$ determined by 
$\delta(\Delta V)=\delta(\lambda_{\rm 0,~core})+\delta(\lambda_{\rm 0,~broad})$, where $\lambda_{\rm 0,~core}$ and 
$\lambda_{\rm 0,~broad}$ mean the measured central wavelengths of the core and the broad \oiii components, and 
$\delta(\lambda_{\rm 0,~core})$ and $\delta(\lambda_{\rm 0,~broad})$ are the corresponding uncertainties of the 
central wavelengths. Then, the following three criteria are applied to collect quasars with reliable broad 
blue-shifted \oiii components. First, the measured parameters of broad H$\beta$ are at least 5 times larger than 
their corresponding uncertainties. Second, the measured line parameters of both core and broad \oiii components 
are at least 5 times larger than their corresponding uncertainties. Third, the shifted velocities $\Delta V$ are 
at least 5 times larger than their corresponding uncertainties: $\Delta V>5\times\delta(\Delta V)$. Based on the 
three criteria, in our final main sample, there are 535 quasars with reliable broad blue-shifted \oiii 
components. Best fitted results to the emission lines in all the 535 quasars can be downloaded\footnote{The 
access code is xip3. There are three files included in the compressed file "Blue\_O3.tar.gz". 
One file "par\_all\_tex.list" includes the necessary parameters of the 535 quasars. One PDF file 
"all\_spec.pdf" (13M) shows the results on the determined stellar lights in the spectra of the 535 SDSS 
quasars with reliable broad blue-shifted \oiii components. Among the 535 SDSS quasars, there are 26 SDSS quasars 
of which spectra include apparent contributions of stellar lights. In the 26 panels with reliable stellar lights, 
line styles have the same meanings as those shown in the left panel of Figure~1. In the other panels, only the 
full SDSS spectra are shown. One PDF file "all\_line.pdf" (21M) shows the results on the corresponding best 
fitted results to the emission lines around H$\beta$ in the line spectra. And the line spectra of the 26 quasars 
are determined by subtractions of the stellar lights from the observed SDSS spectra. There are 17 pages in each 
PDF file, 32 panels per page.} from \url{https://pan.baidu.com/s/1gG85QpuXBDfa4NRb5dQHyg}. The measured line 
parameters of the 535 quasars have been listed in Table~1. And the distribution of final determined 
$\chi^2$ for the best fitted results to the emission lines has been also shown in the left panel of 
Fig.~\ref{dis2} for the 535 quasars. 

\begin{figure*}
\centering\includegraphics[width = 18cm,height=6cm]{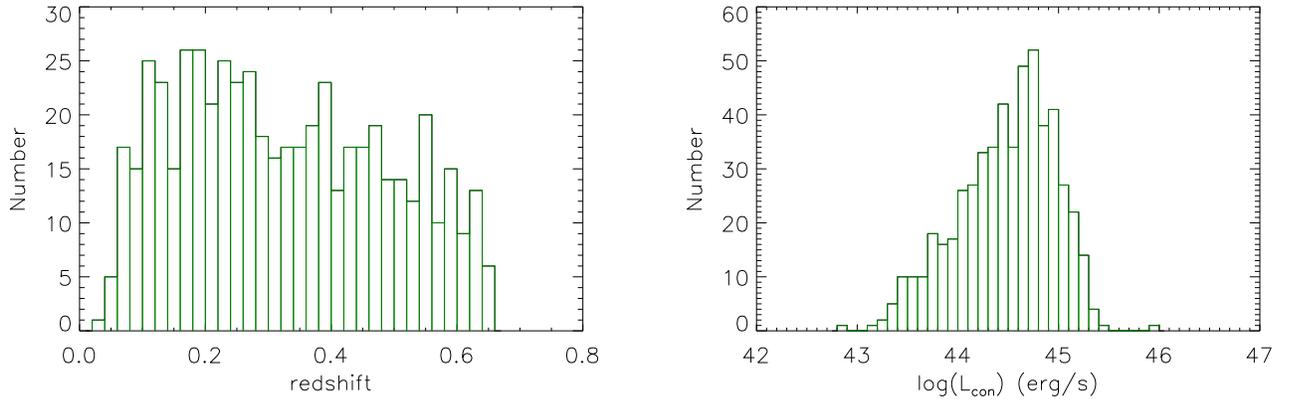}
\caption{Distributions of redshift (left panel) and continuum luminosity at 5100\AA~ (right panel) of the 
535 quasars in our final sample.
}
\label{dis}
\end{figure*}

    Meanwhile, right panel of Fig.~\ref{dis2} shows distributions of the measured line widths of the broad and core 
\oiii components of the collected 535 quasars. In the manuscript, we have accepted second moments as the line 
widths of the core and broad \oiii components. And the maximum line width is about 507${\rm km/s}$ of the core 
\oiii components, but the maximum line width is about 1981${\rm km/s}$ of the broad \oiii components. Fig.~\ref{dis} 
shows distributions of redshift and continuum luminosity at 5100\AA~ ($L_{\rm con}=\lambda L_{\rm 5100\text{\AA}}$) 
of the 535 quasars. The continuum luminosity $L_{\rm con}$ is calculated based on the continuum flux at rest 
wavelength of 5100\AA\ through the determined power law function to describe the AGN continuum emissions underneath 
broad H$\beta$ (such as the component shown as double-dot-dashed red line in the right panel of Fig.~\ref{line}), 
after subtractions of necessary stellar lights. The mean redshift and continuum luminosity are about 0.323 
and $3.02\times10^{44}{\rm erg/s}$, respectively.

    Before proceeding further, there is one point we should note. In some quasars, the broad \oiii components have 
line width $\sigma$ around 1000${\rm km/s}$, such as the line width about $\sigma\sim800{\rm km/s}$ of the broad 
blue-shifted \oiii components in the SDSS 0856-52339-0050 shown in the right panel of Fig.~\ref{line}, which are very 
large values for narrow emission lines, leading to the question whether should the determined broad \oiii components 
be actually as part of broad H$\beta$ from normal BLRs? We rejected the possibility by the following consideration. 
We have checked the low-redshift quasars with $z<0.35$ and with reliable broad \oiii components. If the broad \oiii 
components were part of broad H$\beta$, then similar and corresponding stronger features could be found in the red 
wings of broad H$\alpha$. However, none low redshift quasars have shown such strong features in the red wings of 
broad H$\alpha$. Thus, in the manuscript, we safely accepted that the determined broad \oiii components are truly 
from \oiii emission regions.

\section{Main Results}
\begin{figure*}
\centering\includegraphics[width = 16cm,height=12cm]{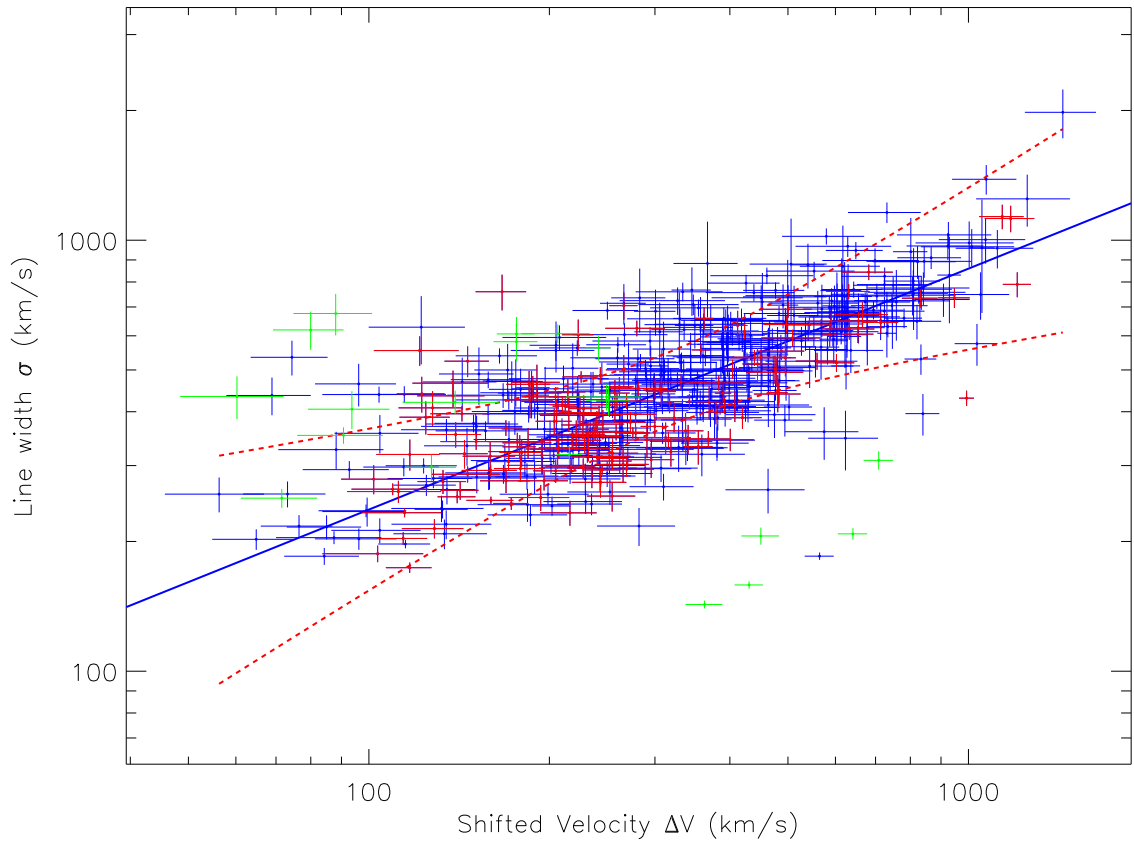}
\caption{On the correlation between $\Delta V$ and line width $\sigma$ of the broad \oiii components in the 
quasars. Blue dots plus error bars show the results for all the 535 quasars, red dots plus error bars 
show the results for the 160 high-quality quasars. Solid blue line and dashed red lines represent the best 
fitted results for all the 535 quasars, and the corresponding $5\sigma$ confidence bands, respectively. Green 
dots plus error bars represent the results for reliable broad red-shifted \oiii components in the 20 quasars.
}
\label{o3}
\end{figure*}

    Based on the well measured line parameters, properties of the broad blue-shifted \oiii components 
can be well checked in the 535 quasars. Fig.~\ref{o3} shows the correlation between blue-shifted 
velocity $\Delta V$ and line width $\sigma$ of the broad blue-shifted \oiii components. A strong linear 
correlation can be confirmed with Spearman Rank correlation coefficient of 0.75 with $P_{null}<10^{-20}$. 
Under considering uncertainties in both coordinates, through the Least Trimmed Squares (LTS) robust 
technique discussed in \citet{cm13}, the correlation can be well described by
\begin{equation}
\log(\frac{\sigma}{\rm km/s})=(1.29\pm0.07)+(0.54\pm0.03)\times\log(\frac{\Delta V}{\rm km/s})
\end{equation},
with rms scatter of about 0.096\ in the space of $\log(\sigma)$ versus $\log(\Delta V)$. In order to ensure 
the strong linear correlation without effects of quality of measured parameters, among the 535 quasars, 
the 160 quasars with parameters (line width and shift velocity) at least 10 times larger than their 
corresponding uncertainties are applied to show the correlation again in Fig~\ref{o3} as red dots. 
The strong linear correlation can be re-confirmed with Spearman Rank correlation coefficient of 0.64 
with $P_{null}\sim10^{-20}$ for the 160 high-quality quasars, with the corresponding rms scatter of about 0.082.

     It is very interesting that it is so-far the first report on so strong linear correlation between $\Delta V$ 
and $\sigma$ in broad blue-shifted \oiii components. The strong linear correlation not similar as the weak trends 
in previous references (such as the result in \citet{eu17, so18}) are mainly due to the following two points. On the 
one hand, the line parameters are measured from the well determined Gaussian broad \oiii components, not similar as 
the asymmetric parameters previously determined from the full \oiii lines. On the other hand, quasars rather than 
type-2 AGN (or hidden type-1 AGN) are considered, leading to much wide parameter range of the shifted velocities of 
broad blue-shifted \oiii components. Based on the strong linear correlation shown in Fig~\ref{o3}, the following 
points have been considered.

    First and foremost, whether the strong linear correlation can be treated as strong evidence for  outflows 
related to central engine but not from local flows in NLRs?  Actually, local flows in NLRs could lead to similar 
detection rates for broad blue-shifted \oiii components and for broad red-shifted \oiii components in quasars. 
However, based on the similar criteria to collect quasars with reliable broad blue-shifted broad \oiii components, 
the criteria are applied to collect quasars with reliable broad red-shifted \oiii components: 
$P_{\rm H\beta}>5\times\delta(P_{\rm H\beta})$, $P_{\rm core}>5\times\delta(P_{\rm core})$, 
$P_{\rm broad}>5\times~\delta(P_{\rm broad})$ and 
$\Delta V=\lambda_{\rm 0,~broad}-\lambda_{\rm 0,~core}>5\times(\delta(\lambda_{\rm 0,~core})+\delta(\lambda_{\rm 0,~broad}))$ 
where $P$ and $\delta(P)$ represent the measured parameters and the corresponding uncertainties, the suffixes of 
"H$\beta$", "core" and "broad" mean the parameters and the uncertainties for the broad H$\beta$, the core \oiii and 
the broad \oiii components, then there are only 20 quasars with reliable broad red-shifted \oiii components. The 
much different detection rates strongly and naturally indicate that the broad blue-shifted \oiii components are 
due to outflows related to central engine rather than due to local flows in NLRs in quasars. Furthermore, for the 
broad red-shifted \oiii components, the correlation between $\Delta V$ and $\sigma$ is very weak with 
correlation coefficient of about -0.28. Besides the much different detection rates, the much different 
correlations between $\Delta V$ and $\sigma$ for the broad blue-shifted and the broad red-shifted \oiii 
components can also strongly indicate the broad blue-shifted \oiii components are due to outflows related to central engine 
rather than due to local flows in NLRs.

   Besides, whether the large shifted velocities of broad blue-shifted \oiii components can be expected by 
outflows related to central engine? As the shown results in Fig.~\ref{o3}, the mean shifted velocity of the 
broad blue-shifted \oiii components of the 535 quasars is about $374{\rm km/s}$ with minimum value about 
$56{\rm km/s}$ and maximum value about $1440{\rm km/s}$. The strong dependence of radial velocities of outflowing 
clouds on distance to central region clearly show that radial velocities on scale of kpcs are about several 
hundreds of kilometers per second (see the discussed results in \citet{kz11, kp15, tm15}), similar as the 
values shown in Fig.~\ref{o3}. Thus, the large shifted velocities are reasonable under the assumption of 
AGN-feedback driven outflows.

   Last but not the least, whether AGN-feedback driven outflows can be applied to explain the strong 
correlation shown in Fig~\ref{o3} for the broad blue-shifted \oiii components in quasars? Under the framework 
of AGN-feedback driven outflows, on scale of kpcs, radial velocities do strongly depend on distance $R$ to 
central regions, $\Delta V\propto R^{-\alpha}$ with $\alpha>0$ (such as the results in \citet{kz11}). Moreover, 
if we accepted that the line width of the broad blue-shifted \oiii components were due to gravitational 
potential of central BH because of the emission regions nearer to central BH, we could have 
$\sigma^2\propto R^{-1}$. Therefore, we can expect the strong correlation between $\Delta V$ and $\sigma$ 
shown in Fig.~\ref{o3}: $\sigma\propto(\Delta V)^{0.5}$, if $\alpha\sim1$ accepted. Unfortunately, we can 
not find an exact mathematical solution to the results shown in Fig.~\ref{o3} based on the AGN-feedback 
driven outflow model, due to much complicated initial physical conditions.

\section{Discussions}

\begin{figure*}
\centering\includegraphics[width = 18cm,height=6cm]{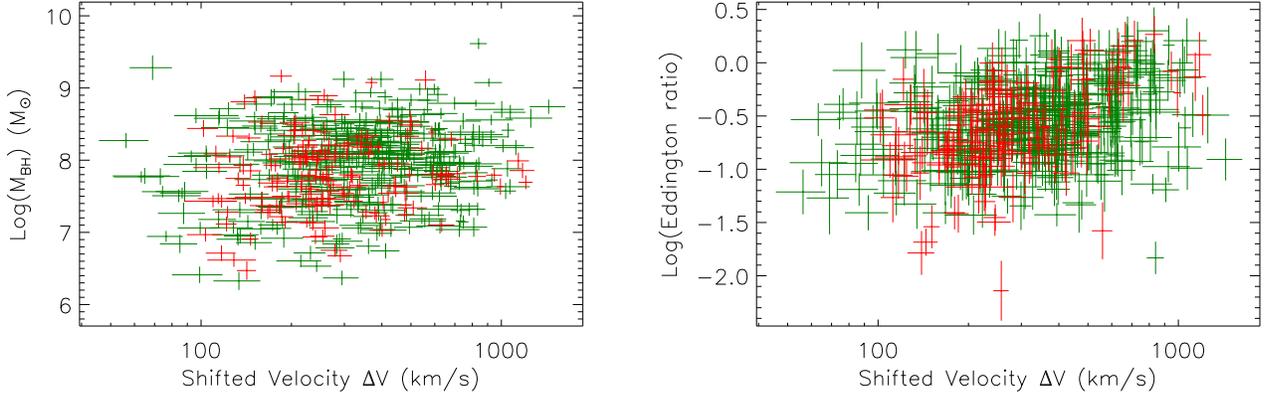}
\caption{On the correlation between BH mass and  $\Delta V$ (left panel) and between Eddington ratio and 
$\Delta V$ (right panel). In each panel, Dots plus error bars are for all the 535 quasars, red dots plus 
error bars are for the 160 high-quality quasars.}
\label{mdm}
\end{figure*}

\begin{figure}
\centering\includegraphics[width = 9cm,height=6cm]{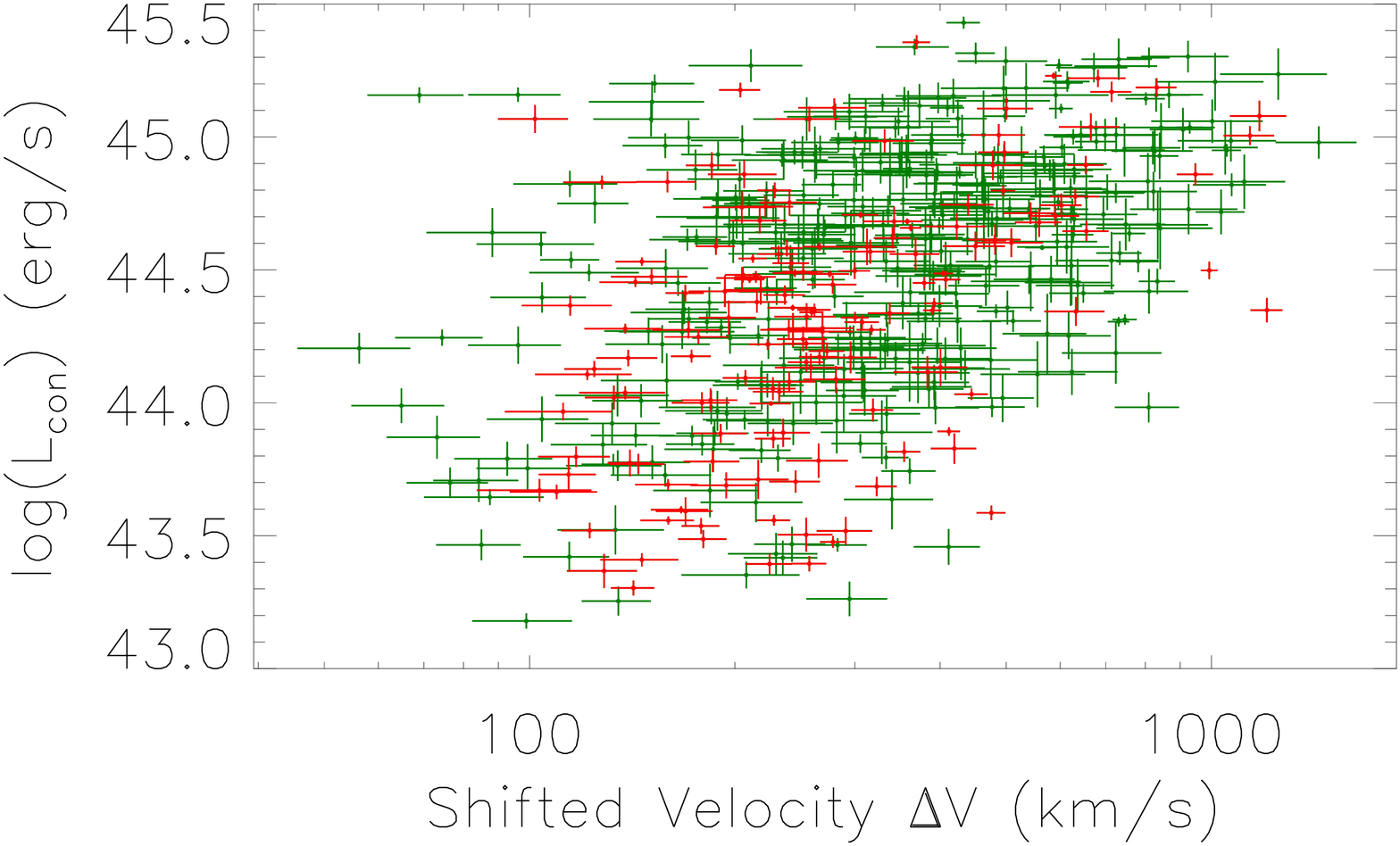}
\caption{On the correlation between continuum luminosity and $\Delta V$. Dots plus error bars are 
	for all the 535 quasars, red dots plus error bars are for the 160 high-quality quasars.}
\label{lum}
\end{figure}

\begin{figure}
\centering\includegraphics[width = 9cm,height=6cm]{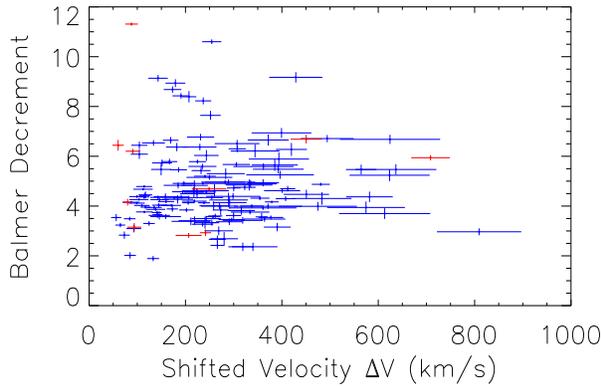}
\caption{On the dependence of $\Delta V$ on Balmer decrements, symbols in blue and in red are for the 
quasars with broad blue-shifted \oiii components, and for the quasars with broad red-shifted \oiii 
components, respectively.}
\label{bd}
\end{figure}

\begin{figure}
\centering\includegraphics[width = 8cm,height=5cm]{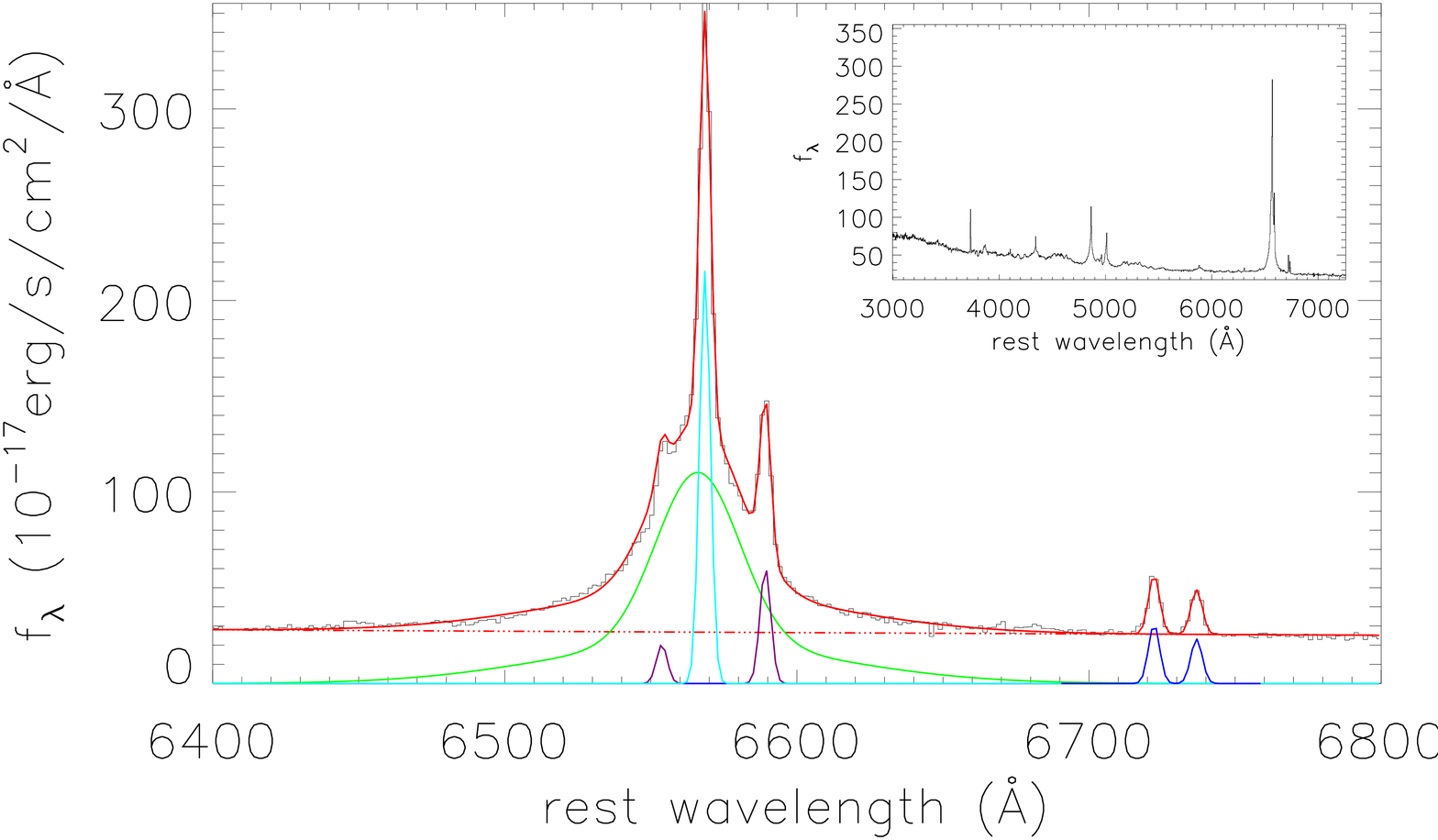}
\caption{An example on the best fitted results to the emission lines around H$\alpha$ in the quasar SDSS 
1846-54173-0426. In the figure, solid line in black and solid line in red show the line spectrum and the best fitted 
results, double-dot-dashed red line shows the determined power law continuum emissions, solid green line shows the 
determined broad H$\alpha$, solid purple lines and in blue show the determined \nii and \sii doublets, solid cyan 
line shows the determined narrow H$\alpha$, respectively. In the top-right corner, the full spectrum with few 
contributions of stellar lights has been shown. }
\label{msp}
\end{figure}

\begin{figure*}
\centering\includegraphics[width = 18cm,height=6cm]{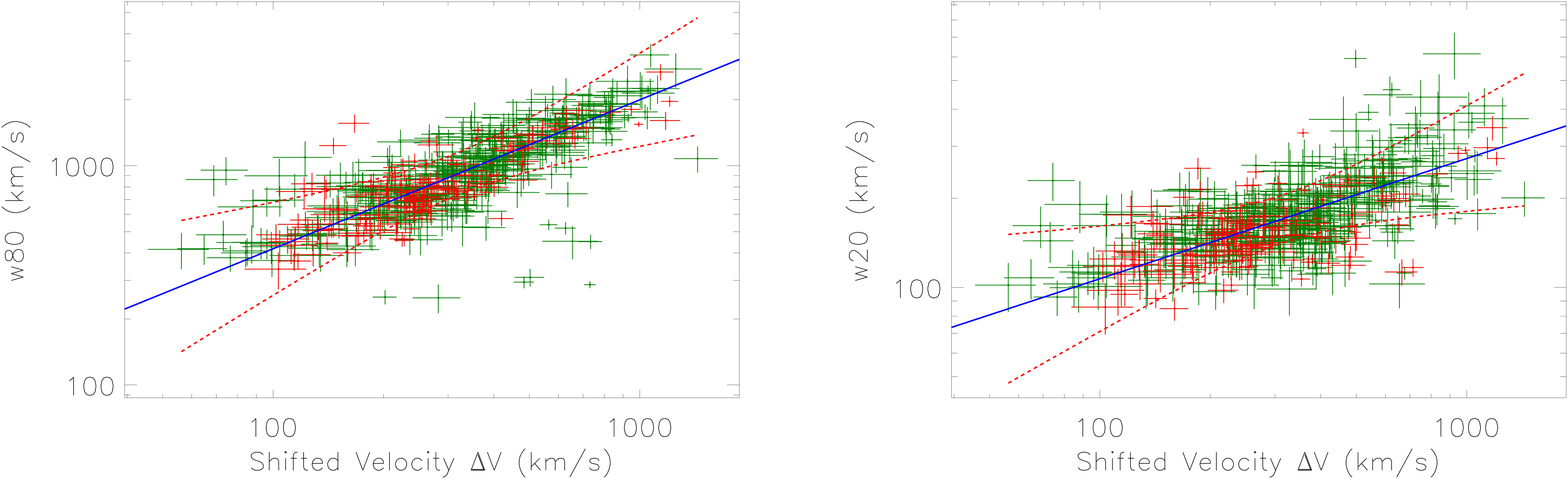}
\caption{Left panel shows the correlation between $w80$ and $\Delta V$. Right panel shows the correlation between 
$w20$ and $\Delta V$. In each panel, solid blue line shows the best fitted results, dashed red lines show 
the corresponding $5\sigma$ confidence bands, respectively. In each panel, dots are for 
all the 535 quasars, red dots plus error bars are for the 160 high-quality quasars.}
\label{w80}
\end{figure*}

   In the section, some further discussions have been shown on the dependence of the strong linear correlation shown 
in Fig.~\ref{o3} on the other physical parameters, such as Eddington ratio, BH mass, etc.. The virial BH mass 
$M_{\rm BH}$ can be well determined by the line parameters of broad H$\beta$ \citep{gh05, vp06, sh11} under the 
Virialization assumption to broad line emission clouds \citep{pe04},
\begin{equation}
\frac{M_{\rm BH}}{\rm M_\odot}=3.6\times10^6\times(\frac{L_{\rm H\beta}}{\rm 10^{42}erg/s})^{0.56}
	\times(\frac{FWHM_{\rm H\beta}}{\rm 1000km/s})^2
\end{equation}, 
where $L_{\rm H\beta}$ and $FWHM_{\rm H\beta}$ represent the line luminosity and full width at half maximum of 
broad H$\beta$, respectively. The Eddington ratio $\dot{M}_{\rm Edd}$ can be applied to trace intrinsic 
accretion rate and estimated by 
\begin{equation}
\dot{M}_{\rm Edd}=\frac{10\times L_{\rm 5100}}{1.4\times10^{38}M_{\rm BH}/{\rm M_{\odot}}}
\end{equation}.
Here, the accepted optical bolometric correction $L_{\rm bol}\sim10\times L_{\rm 5100}$ is mainly from the 
statistical properties of spectral energy distributions of a sample of low redshift quasars discussed in \citet{rg06} 
and from the more recent discussed results in \citet{nh20} based on theoretical calculations. Moreover, \citet{db20} 
have shown that the optical bolometric corrections appear to be fairly constant. Therefore, as commonly applied 
in the literature \citep{kas00, sh11}, we safely accepted the optical bolometric correction 
$L_{\rm bol}\sim10\times L_{\rm 5100}$.

     Based on the calculated $M_{\rm BH}$ and $\dot{M}_{\rm Edd}$, the Spearman Rank correlation coefficients 
are about 0.15 ($P_{\rm null}\sim4\times10^{-4}$) and 0.36 ($P_{\rm null}\sim1.3\times10^{-17}$) for the correlation 
between $M_{\rm BH}$ and $\Delta V$ and for the correlation between $\dot{M}_{\rm Edd}$ and $\Delta V$, respectively, 
shown in Fig.~\ref{mdm}. It is clear that rather than central BH masses, accretion rates have more important 
effects on the strong linear correlation shown in Fig.~\ref{o3}: more powerful accretion rates lead to more stronger 
blue-shifted \oiii components. Meanwhile, besides the moderate dependence of Eddington ratios on the strong correlation 
between $\Delta V$ and $\sigma$, Fig.~\ref{lum} shows the correlation between $\Delta V$ and continuum luminosity at 
5100\AA. The correlation is stronger, with Spearman Rank correlation coefficient of about 0.48 for all the 535 quasars 
and of about 0.49 for the 160 high-quality quasars. Therefore, rather than BH masses, Eddington ratios and continuum 
luminosities have more important roles on properties of broad blue-shifted \oiii components in quasars.

   Moreover, we have checked effects of extinctions traced by Balmer decrements (flux ratios of narrow H$\alpha$ to 
narrow H$\beta$). Fig.~\ref{bd} shows the correlation between Balmer decrement and $\Delta V$ of the 151 low-redshift 
quasars with reliable broad blue-shifted \oiii components and with reliable narrow H$\alpha$ and narrow H$\beta$ 
emission lines, and of the 10 low-redshift quasars with reliable broad red-shifted \oiii components and with reliable 
narrow H$\alpha$ and narrow H$\beta$ emission lines. The Spearman Rank correlation coefficients are about 0.12 and -0.07 
for the quasars with broad blue-shifted and with broad red-shifted \oiii components, respectively. The mean 
Balmer decrements are about 4.71 and 4.77 for the quasars with broad blue-shifted and with broad red-shifted \oiii 
components, respectively. Therefore, there are few effects of extinctions on the linear correlation between $\Delta V$ and 
$\sigma$. And moreover, in spite of less number of quasars with broad red-shifted \oiii components, there should be no 
different effects of extinctions on broad blue-shifted and broad red-shifted \oiii components. Here, the emission lines 
around H$\alpha$ are measured as what we have done to measure emission lines around H$\beta$. There are two (or more, 
if necessary) Gaussian components applied to describe broad H$\alpha$, one narrow Gaussian component applied to describe 
narrow H$\alpha$, two Gaussian components applied to describe [N~{\sc ii}]$\lambda6548,~6583$\AA~ doublet, two Gaussian 
components applied to describe [S~{\sc ii}]$\lambda6716,~6732$\AA~ doublet and one power law component applied to 
describe continuum emissions. As an example, Fig.~\ref{msp} shows the best fitted results 
to the emission lines around H$\alpha$ in the quasar SDSS 1846-54173-0426.

   Furthermore, as more recent discussed properties of \oiii emissions in \citet{ma13, ha14, km16, wf20}, the parameter 
$w80$ has been well applied to parameterise the velocity width of asymmetric \oiii emission lines, which refers to 
the velocity width that encloses 80\% of the total \oiii flux. Here, two parameters of $w80$ and $w20$ of the \oiii 
emission lines are calculated in the SDSS quasars, where $w20$ refers to the velocity width that encloses 20\% of the 
total \oiii flux. Then, rather than the line width of broad \oiii components applied , Fig.~\ref{w80} shows the 
correlation between $w80$ ($w20$) and $\Delta V$, which can be well described by 
\begin{equation}
\begin{split}
\log(\frac{w80}{\rm km/s})&=(1.26\pm0.04)+(0.68\pm0.02)\times\log(\frac{\Delta V}{\rm km/s}) \\
\log(\frac{w20}{\rm km/s})&=(1.22\pm0.03)+(0.41\pm0.02)\times\log(\frac{\Delta V}{\rm km/s})
\end{split}
\end{equation},
through the LTS technique. It is apparent that there are also two linear correlations, but the Spearman Rank 
correlation coefficient is about 0.79 ($P_{\rm null}<10^{-20}$) for the correlation between $w80$ and $\Delta V$, 
larger than the coefficient of 0.63 for the correlation between $w20$ and $\Delta V$. The stronger correlation on 
the parameter of $w80$ indicates broad blue-shifted \oiii component rather than the core \oiii components have 
more important contributions to the correlation between $\sigma$ and $\Delta V$ shown in Fig.~\ref{o3}.

\section{conclusions}

     Finally, we give our main conclusions as follows. Based on a large sample of 535 SDSS quasar with reliable 
broad blue-shifted \oiii components, a strong linear correlation with Spearman Rank correlation coefficient 0.75 
can be clearly confirmed between shifted velocity and line width of the broad blue-shifted \oiii components, 
which can be explained under the assumption of 
AGN-feedback driven outflows. Meanwhile, through similar criteria, there are only 20 SDSS quasars with reliable 
broad red-shifted \oiii components, and there is no positive correlation between shifted velocity and line width 
of the broad red-shifted \oiii components, providing strong evidence against the blue-shifted \oiii components related to 
local flows in NLRs. Therefore, strong broad blue-shifted \oiii components can be treated as better indicator 
of outflows related to central engine in AGN. Moreover, stronger dependence of Eddington ratios and continuum 
luminosities can be found on the correlation between $\Delta V$ and $\sigma$ of the broad blue-shifted \oiii 
components in quasars, indicating Eddington ratios and continuum luminosities have more important roles on properties 
of blue-shifted \oiii components in quasars.

\section*{Acknowledgements}
Zhang gratefully acknowledges the anonymous referee for giving us constructive comments and suggestions 
greatly improving our paper. Zhang acknowledges the kind financial support from Nanjing Normal University and kind 
support from the Chinese grant NSFC-11973029. This paper has made use of the data from the SDSS projects. The SDSS-III 
web site is http://www.sdss3.org/. SDSS-III is managed by the Astrophysical Research Consortium for the Participating 
Institutions of the SDSS-III Collaboration.

\begin{table*}
\caption{Parameters of the 535 SDSS quasars with broad blue-shifted \oiii components}
\begin{tabular}{ccccccccc}
\hline\hline
pmf & z & $\Delta V$   & $\sigma$  & $\log(L_{\rm con})$ & $\log(M_{\rm BH})$ & $\log(\dot{M}_{\rm Edd})$ & $w80$ & $w20$\\
&   & ${\rm km/s}$ & ${\rm km/s}$ & ${\rm erg/s}$    & ${\rm M_\odot}$  &     &  ${\rm km/s}$ & ${\rm km/s}$ \\
\hline
0274-51913-0388   &   0.256   &   392$\pm$51   &   478$\pm$28   &   44.07$\pm$0.06   &   7.75   &   0.111   &   870   &   158   \\
0276-51909-0038   &   0.187   &   207$\pm$15   &   330$\pm$19   &   44.09$\pm$0.02   &   7.39   &   0.086   &   660   &   132   \\
0282-51658-0138   &   0.268   &   512$\pm$50   &   669$\pm$75   &   44.30$\pm$0.04   &   8.01   &   0.094   &   1385   &   229   \\
0285-51930-0280   &   0.259   &   238$\pm$17   &   364$\pm$22   &   44.58$\pm$0.03   &   8.40   &   0.087   &   515   &   133   \\
0291-51928-0456   &   0.089   &   208$\pm$41   &   595$\pm$40   &   43.35$\pm$0.05   &   7.63   &   0.092   &   1035   &   117   \\
0294-51986-0528   &   0.275   &   439$\pm$74   &   756$\pm$37   &   44.61$\pm$0.06   &   8.33   &   0.099   &   1491   &   263   \\
0299-51671-0098   &   0.325   &   1117$\pm$165   &   958$\pm$98   &   44.83$\pm$0.10   &   8.17   &   0.104   &   2246   &   410   \\
0299-51671-0251   &   0.385   &   389$\pm$36   &   573$\pm$83   &   44.94$\pm$0.03   &   8.20   &   0.098   &   1304   &   185   \\
0301-51641-0151   &   0.163   &   223$\pm$42   &   421$\pm$25   &   43.93$\pm$0.05   &   7.42   &   0.097   &   908   &   141   \\
0309-51994-0116   &   0.132   &   144$\pm$13   &   320$\pm$24   &   43.76$\pm$0.03   &   8.30   &   0.063   &   515   &   114   \\
\hline\hline
\end{tabular}\\
Notice: Col(1) shows the information of SDSS plate-mjd-fiberid, Col(2) shows the information of redshift, 
Col(3) shows the shifted velocity of the broad \oiii components relative to the core \oiii components, 
Col(4) shows the line width of the broad \oiii components, Col(5) shows the information of continuum luminosity, 
Col(6) shows the information of virial BH masses, Col(7) shows the information of Eddington ratios, 
Col(8) and Col(9) show the information of $w80$ and $w20$. \\
Here, parameters of only 10 of the 535 quasars are listed in the table. The full table can be downloaded from 
\url{https://pan.baidu.com/s/1gG85QpuXBDfa4NRb5dQHyg} with access code xip3.
\end{table*}

\label{lastpage}

\begin{thebibliography}{   }
\bibitem[\protect\citeauthoryear{Abazajian et al.}{2003}]{ab03}
Abazajian, K.; Adelman-McCarthy, J. K.; Agueros, M. A.; et al., 2003, AJ, 126, 2081
\bibitem[\protect\citeauthoryear{Alam et al.}{2015}]{al15}
Alam, S., et al., 2015, ApJS, 219, 12
\bibitem[\protect\citeauthoryear{Boroson}{2005}]{tb05}
Boroson, T., 2005, AJ, 130, 381
\bibitem[\protect\citeauthoryear{Bruzual \& Charlot}{2003}]{bc03}
Bruzual, G., \& Charlot, S. 2003, MNRAS, 344, 1000
\bibitem[\protect\citeauthoryear{Cappellari et al.}{2013}]{cm13}
Cappellari, M.; Scott, N.; Alatalo, K.; et al., 2013, MNRAS, 432, 1709
\bibitem[\protect\citeauthoryear{Cid Fernandes et al.}{2005}]{cm05}
Cid Fernandes, R., Mateus, A., Sodre, L., Stasinska, G., Gomes, J. M., 2005, MNRAS, 358, 363
\bibitem[\protect\citeauthoryear{Cheung et al.}{2016}]{cb16}
Cheung, E.; Bundy, K.; Cappellari, M.; et al., 2016, Natur, 533, 504
\bibitem[\protect\citeauthoryear{Crenshaw et al.}{2003}]{ck03}
Crenshaw, D. M.; Kraemer, S. B.; George, I. M., 2013, ARA\&A, 41, 117
\bibitem[\protect\citeauthoryear{Dimitrijevic et al.}{2007}]{dk07}
Dimitrijevic, M. S.; Kovacevic, J.; Popovic, L. C.; Dacic, M.; Ilic, D., 2007, MNRAS, 374, 1181
\bibitem[\protect\citeauthoryear{Duras et al.}{2020}]{db20}
Duras, F.; Bongiorno, A.; Ricci, F.; et al., 2020, A\&A, 636, 73
\bibitem[\protect\citeauthoryear{Eun et al.}{2017}]{eu17}
Eun, D.; Woo, J.-H.; Bae, H.-J., 2017, ApJ, 842, 5
\bibitem[\protect\citeauthoryear{Fabian}{2012}]{fa12}
Fabian, A. C., 2012, ARA\&A, 50, 455
\bibitem[\protect\citeauthoryear{Ferrarese \& Merritt}{2000}]{fm00}
Ferrarese, F. \& Merritt, D., 2000, ApJL, 539, L9
\bibitem[\protect\citeauthoryear{Foreman-Mackey et al.}{2013}]{fh13}
Foreman-Mackey, D.; Hogg, D. W.; Lang, D.; Goodman, J., 2016, PASP, 125, 306
\bibitem[\protect\citeauthoryear{Ganguly et al.}{2007}]{gb07}
Ganguly, R.; Brotherton, M. S.; Cales, S.; et al., 2007, ApJ, 665, 990
\bibitem[\protect\citeauthoryear{Gebhardt et al.}{2000}]{geb00}
Gebhardt, K.; Bender, R.; Bower, G.; et al., 2000, ApJL, 539, L13
\bibitem[\protect\citeauthoryear{Greene \& Ho}{2005}]{gh05}
Green, J. E., Ho, L. C., 2005, ApJ, 630, 122
\bibitem[\protect\citeauthoryear{Gupta et al.}{2005}]{gs05}
Gupta, Neeraj; Srianand, R.; Saikia, D. J.. 2005, MNRAS, 361, 451
\bibitem[\protect\citeauthoryear{Harrison et al.}{2014}]{ha14}
Harrison, C. M.; Alexander, D. M.; Mullaney, J. R.; Swinbank, A. M., 2014, MNRAS, 441, 3306
\bibitem[\protect\citeauthoryear{Holt et al.}{2008}]{ht08}
Holt, J.; Tadhunter, C. N.; Morganti, R., 2008, MNRAS, 387, 639
\bibitem[\protect\citeauthoryear{Kaspi et al.}{2000}]{kas00}
Kaspi, S.; Smith, P. S.; Netzer, H.; et al., 2000, ApJ, 533, 631
\bibitem[\protect\citeauthoryear{Kauffmann et al.}{2003}]{ka03}
Kauffmann, G., et al. 2003, MNRAS, 346, 1055
\bibitem[\protect\citeauthoryear{Kakkad et al.}{2016}]{km16}
Kakkad, D.; Mainieri, V.; Padovani, P.; et al., 2016, A\&A, 592, 148
\bibitem[\protect\citeauthoryear{King et al.}{2011}]{kz11}
King, A. R.; Zubovas, K.; Power, C., 2011, MNRAS Letter, 415, 6
\bibitem[\protect\citeauthoryear{King \& Pounds}{2015}]{kp15}
King, A.; Pounds, K., 2015, ARA\&A, 53, 115
\bibitem[\protect\citeauthoryear{Kormendy \& Ho}{2013}]{kh13}
Kormendy, J. \& Ho, L. C., 2013, ARA\&A, 51, 511
\bibitem[\protect\citeauthoryear{Kovacevic et al.}{2010}]{kp10}
Kovacevic, J., Popovic, L. C., Dimitrijevic, M. S., 2010, ApJS, 189, 15
\bibitem[\protect\citeauthoryear{Martin-Navarro et al.}{2018}]{mb18}
Martin-Navarro, I.; Brodie, J. P.; Romanowsky, A. J.; Ruiz-Lara, T.; van de Ven, Glenn, 2018, Natur, 553, 307
\bibitem[\protect\citeauthoryear{Mullaney et al.}{2013}]{ma13}
Mullaney, J. R.; Alexander, D. M.; Fine, S.; et al., 2013, MNRAS, 433, 622
\bibitem[\protect\citeauthoryear{Netzer}{2020}]{nh20}
Netzer, H., 2020, MNRAS, 488, 5185
\bibitem[\protect\citeauthoryear{Page et al.}{2012}]{ps12}
Page, M. J.; Symeonidis, M.; Vieira, J. D.; et al., 2012, Natur, 485, 213
\bibitem[\protect\citeauthoryear{Perna et al.}{2015}]{pb15}
Perna, M.; Brusa, M.; Cresci, G.; et al., 2015, A\&A, 574, 82
\bibitem[\protect\citeauthoryear{Peterson et al.}{2004}]{pe04}
Peterson, B. M.; Ferrarese, L.; Gilbert, K. M., et al., 2004, ApJ, 613, 682
\bibitem[\protect\citeauthoryear{Rakshit et al.}{2017}]{ra17}
Rakshit, S.; Stalin, C. S.; Chand, H.; Zhang, X. G., 2017, ApJS, 229, 39
\bibitem[\protect\citeauthoryear{Shen et al.}{2011}]{sh11}
Shen, Y.; Richards, G. T.; Strauss, M. A; et al., 2011, ApJS, 194, 45
\bibitem[\protect\citeauthoryear{Sergeev et al.}{1997}]{sp97}
Sergeev, S. G.; Pronik, V. I.; Malkov, Y. F.; Chuvaev, K. K., 1997, A\&A, 320, 405
\bibitem[\protect\citeauthoryear{Richards et al.}{2002}]{rg02}
Richards, G. T., et al., 2002, AJ, 123, 2945
\bibitem[\protect\citeauthoryear{Richards et al.}{2006}]{rg06}
Richards, G. T.; Lacy, M.; Storrie-Lombardi, L. J.; et al., 2006, ApJS, 166, 470 
\bibitem[\protect\citeauthoryear{Ross et al.}{2012}]{ro12}
Ross, N. P., et al., 2012, ApJS, 199, 3
\bibitem[\protect\citeauthoryear{Schmidt et al.}{2018}]{so18}
Schmidt, E. O.; Oio, G. A.; Ferreiro, D.; Vega, L.; Weidmann, W., 2018, A\&A, 615, 13
\bibitem[\protect\citeauthoryear{Tadhunter et al.}{2001}]{tw01}
Tadhunter, C.; Wills, K.; Morganti, R.; Oosterloo, T.; Dickson, R., 2001, MNRAS, 327, 227
\bibitem[\protect\citeauthoryear{Tombesi et al.}{2015}]{tm15}
Tombesi, F.; Meléndez, M.; Veilleux, S.; et al., 2015, Natur, 519, 436
\bibitem[\protect\citeauthoryear{Veilleux et al.}{2005}]{vc05}
Veilleux, S.; Cecil, G.; Bland-Hawthorn, J., 2005, ARA\&A, 43, 769
\bibitem[\protect\citeauthoryear{Vestergaard \& Peterson}{2006}]{vp06}
Vestergaard, M., Peterson, B. M. 2006, ApJ, 641, 689
\bibitem[\protect\citeauthoryear{Wylezalek et al.}{2020}]{wf20}
Wylezalek, D.; Flores, A. M.; Zakamska, N. L.; Greene, J. E.; Riffel, R. A., et al., 2020, MNRAS, 492, 4680
\bibitem[\protect\citeauthoryear{Woo et al.}{2016}]{wo16}
Woo, J.-H.; Bae, H.-J.; Son, D.; Karouzos, M., 2016, ApJ, 817, 108
\bibitem[\protect\citeauthoryear{Zakamska et al.}{2016}]{za16}
Zakamska, N. L.; Hamann, F.; Paris, I.; et al., 2016, MNRAS, 459, 3144
\bibitem[\protect\citeauthoryear{Zhang}{2014}]{zh14}
Zhang, X. G., 2014, MNRAS, 438, 557
\bibitem[\protect\citeauthoryear{Zhang et al.}{2016}]{zh16}
Zhang, X. G., Feng, L. L., 2016, MNRAS, 457, 3878
\bibitem[\protect\citeauthoryear{Zhang et al.}{2017}]{zh17}
Zhang, X. G. \& Feng, L. L., 2017, MNRAS, 468, 620
\bibitem[\protect\citeauthoryear{Zhang et al.}{2019}]{zh19}
Zhang, X. G.,; Bao, M., Yuan, Q. R., 2019, MNRAS Letter, 490, 81
\end{thebibliography}
\end{document}